# 車聯網標準安全技術評估分析—探索破解 ECQV 隱式憑證

Abel C. H. Chen
中華電信研究院
資通安全研究所
高級研究員
0000-0003-3628-3033

**摘要**

　　IEEE 1609.2 和 IEEE 1609.2.1 標準中為車聯網(Vehicle-to-everything, V2X)安全通訊，採用和設計了基於橢圓曲線密碼學(Elliptic Curve Cryptography, ECC)的多種安全演算法。為了提高安全憑證管理系統(Security Credential Management System, SCMS)的效能，本研究評估了在不同安全強度下的金鑰產製、金鑰擴展、簽章產製、以及簽章驗證的計算時間。討論基於橢圓曲線 Qu-Vanstone (Elliptic Curve Qu-Vanstone, ECQV)隱式憑證的相關技術，分析未壓縮橢圓曲線點、壓縮橢圓曲線點、顯式憑證、以及隱式憑證的長度。此外，本研究提出了數學模型來證明 ECQV 破解的機率，並提出了避免 ECQV 被破解的建議。

關鍵字：IEEE 1609.2 標準、IEEE 1609.2.1 標準、橢圓曲線數位簽章演算法、橢圓曲線點壓縮、橢圓曲線 Qu-Vanstone 隱式憑證。

## 一、前言

　　為提供安全的車聯網(Vehicle-to-everything, V2X)通訊，IEEE 1609.2 和 IEEE 1609.2.1 標準採用了基於橢圓曲線密碼學(Elliptic Curve Cryptography, ECC)的相關方法，包括橢圓曲線數位簽章演算法(Elliptic Curve Digital Signature Algorithm, ECDSA) [1]和橢圓曲線 Qu-Vanstone (Elliptic Curve Qu-Vanstone, ECQV)隱式憑證方法[2]-[4]。由於車聯網通訊的封包傳輸長度有限，所以需要使用較短的安全協定資料單元(Secure Protocol Data Unit, SPDU)，並且需要簡短的簽名和簡短的憑證。因此，在安全憑證管理系統(Security Credential Management System, SCMS)中，一個可以快速產製和驗證較短簽章的演算法是關鍵的技術。此外，可以結合橢圓曲線(Elliptic Curve, EC)點壓縮方法和 ECQV 隱式憑證方法[5]-[6]獲得較短的憑證。

　　為了評估 IEEE 1609.2 和 IEEE 1609.2.1 標準中的安全技術效能，本研究測量了在不同安全強度下金鑰產製、金鑰擴展、基於橢圓曲線數位簽章演算法的簽章產製和簽章驗證的計算時間。本研究還分析和討論了未壓縮橢圓曲線點、壓縮橢圓曲線點、顯式憑證、以及基於 ECQV 隱式憑證的長度。此外，本研究發現了 ECQV 破解的潛在威脅，並提出了一個數學模型來證明 ECQV 被破解的機率。最後，提出了一些建議，以改善安全憑證管理系統的效能，並避免 ECQV 被破解的潛在威脅。

　　本論文分為七個章節。第二節評估基於橢圓曲線密碼學的金鑰產製和金鑰擴展的效能。第三節說明基於橢圓曲線數位簽章演算法的簽章產製和簽章驗證的計算時間。第四節討論未壓縮橢圓曲線點和壓縮橢圓曲線點的長度。第五節描述 ECQV 的概念並證明 ECQV 破解機率。最後，第六節討論相關建議來提升系統效率和安全，第七節總結本研究的貢獻。

## 二、評估橢圓曲線密碼學

　　本節介紹橢圓曲線密碼學(Elliptic Curve Cryptography, ECC)在金鑰產製和金鑰擴展的作法。公式(1)展示一個橢圓曲線的示例，具有坐標$(x, y)$、係數 $a$ 和 $b$，以及質數 $n$。

在美國國家標準與技術研究院(National Institute of Standards and Technology, NIST)的金鑰產製標準中，定義了基點 $G$、橢圓曲線點加法(EC point addition, ECPA)和橢圓曲線點倍增(EC point doubling, ECPD) [7]-[8]。

$$y^2 = x^3 + ax^2 + b \pmod{n}. \tag{1}$$

### 2.1、金鑰產製

產製 ECC 金鑰對的流程中，首先生成一個隨機整數 $p$ 作為私鑰，然後可以根據基點 $G$ 使用公式(2)得到公鑰 $P$。在公式(2)的計算中，主要根據 $p$ 值執行了橢圓曲線點加法和橢圓曲線點倍增的操作。因此，ECC 金鑰產製主要計算時間花費在計算橢圓曲線點 $P$ [7]-[8]。

$$P = pG. \tag{2}$$

### 2.2、金鑰擴展

金鑰擴展是在車聯網標準中常用的技術，主要結合蝴蝶金鑰擴展(Butterfly Key Expansion, BKE)機制和產製隱式憑證[4]。金鑰擴展的核心想法是產生一個整數 $r$ (可以是隨機整數或加密整數)，將其與原始私鑰 $p$ 結合產製擴展私鑰 $e$ (如公式(3)所示)。此外，可以根據橢圓曲線點 $R$ (即 $R = rG$)使用公式(4)確定擴展的公鑰 $E$。因此，ECC 金鑰擴展主要計算時間花費在計算橢圓曲線點 $R$，金鑰擴展的計算時間與金鑰產製的計算時間類似。

$$e = p + r. \tag{3}$$

$$E = eG = (p+r)G = P + R, \text{where } R = rG. \tag{4}$$

### 2.3、實證結果

為了評估金鑰產製和金鑰擴展的計算時間，本研究採用 Microsoft Windows 10 的筆記型電腦。該筆記型電腦配備了 Intel(R) Core(TM) i7-10510U CPU、8 GB 內存、OpenJDK 18.0.2.1 和 Java Library Bouncy Castle Release 1.72。

本研究根據美國國家標準與技術研究院定義的安全強度[9]，量測金鑰產製和金鑰擴展的計算時間(如表一所示)。例如，基於 NIST P-256 的金鑰產製所需的時間為 12.699 毫秒，而使用相同橢圓曲線進行金鑰擴展的時間為 12.242 毫秒。此外，當使用較大的整數作為私鑰以提高安全強度時，金鑰產製和金鑰擴展的計算時間都會增加。其中，NIST P-256 已被選為 IEEE 1609.2 和 1609.2.1 標準中 SCMS 的主要橢圓曲線之一，因此 NIST P-256 的實驗結果可以作為 V2X 應用程序開發的參考。

表一: 金鑰產製和金鑰擴展的計算時間(單位：毫秒)

| 安全強度 | 橢圓曲線 | 金鑰產製 | 金鑰擴展 |
|---|---|---|---|
| 80 | NIST P-192 | 9.041 | 8.552 |
| 112 | NIST P-224 | 9.345 | 9.052 |
| 128 | NIST P-256 | 12.699 | 12.242 |
| 192 | NIST P-384 | 18.239 | 18.042 |
| 256 | NIST P-521 | 20.167 | 19.504 |

## 三、評估橢圓曲線數位簽章演算法

本節介紹橢圓曲線數位簽章演算法的簽章產製和簽章驗證的作法。其中，以私鑰 $p$ 被選擇用於為消息 $m$ 產製簽章，而公鑰 $P$ 則用於驗證簽章。

### 3.1、簽章產製

為了基於私鑰 $p$ 和消息 $m$ 產製簽章，首先產製一個隨機整數 $r$，並在公式(5)中使用。此外，相應的橢圓曲線點 $R$ (即 $R = rG$)，其座標為 $(R_x, R_y)$，也被納入公式(5)中進行簽章產製。應用公式(5)後，簽章可以生成為 $(R, s)$。因此，ECDSA 簽章產製主要計算時間花費在計算橢圓曲線點 $R$ [1]。

$$s = [(m + pR_x) / r](\mod n). \tag{5}$$

### 3.2、簽章驗證

要驗證簽章 $(R, s)$，可以基於 $s$ 值使用公式(6)、(7)、以及(8)計算三個臨時值 $t_1$、$t_2$、以及 $t_3$。此外，可以根據公鑰 $P$、臨時值 $t_2$、以及臨時值 $t_3$ 使用公式(9)計算橢圓曲線點 $T$。如果橢圓曲線點 $T$ 等於橢圓曲線點 $R$，則簽章驗證成功。公式(10)證明簽章驗證的原理證明[1]。因此，ECDSA 簽章驗證主要計算時間花費在計算橢圓曲線點 $t_2G$ 和橢圓曲線點 $t_3P$，導致與簽章產製相比，簽章驗證的計算時間較長。

$$t_1 = 1/s = [r/(m+pR_x)](\mod n). \tag{6}$$
$$t_2 = mt_1 = m/s = mr/(m+pR_x). \tag{7}$$
$$t_3 = R_x t_1 = R_x/s = R_x r/(m+pR_x). \tag{8}$$
$$T = t_2 G + t_3 P = t_2 G + t_3 pG = (t_2 + t_3 p)G = rG = R. \tag{9}$$
$$t_2 + t_3 p = [mr/(m+pR_x)] + [pR_x r/(m+pR_x)] = r. \tag{10}$$

### 3.3、實證結果

本研究根據美國國家標準與技術研究院定義的安全強度[9]，量測 ECDSA 簽章產製和簽章驗證的計算時間(如表二所示)。其中，實驗環境如 2.3 節所述。基於 NIST P-256 的簽章產製所需的時間為 0.748 毫秒，而使用相同橢圓曲線進行簽章驗證的時間為 0.777 毫秒。這些結果表明，簽章驗證的計算時間比簽章生成的時間更長。此外，當將更大的整數用作私鑰以提高安全強度時，簽章產製和簽章驗證的時間都會增加。根據 IEEE 1609.2 和 1609.2.1 標準，V2X 開發者可以參考使用 NIST P-256 的實驗結果來構建他們的終端設備和 SCMS。

表二: 簽章產製和簽章驗證的計算時間(單位：毫秒)

| 安全強度 | 橢圓曲線 | 簽章產製 | 簽章驗證 |
|---|---|---|---|
| 80 | NIST P-192 | 0.639 | 0.697 |
| 112 | NIST P-224 | 0.731 | 0.750 |
| 128 | NIST P-256 | 0.748 | 0.777 |
| 192 | NIST P-384 | 1.148 | 1.313 |
| 256 | NIST P-521 | 1.376 | 1.382 |

## 四、評估橢圓曲線點壓縮

4.1 節比較未壓縮橢圓曲線點(Uncompressed EC Point)和已壓縮橢圓曲線點(Compressed EC Point)的結構，並且在 4.2 節討論了已壓縮橢圓曲線點的應用。

## 4.1、未壓縮橢圓曲線點和已壓縮橢圓曲線點的結構

為了表示未壓縮橢圓曲線點和已壓縮橢圓曲線點的結構，IEEE 1609.2 標準定義了 EccP256CurvePoint 的格式[5]。該格式包含 CHOICE 用來指出橢圓曲線點的類型，CHOICE 由一個 byte 組成。其中，主要包含 x-only、fill、compressed-y-0、compressed-y-1、以及 uncompressedP256 等類型。由於 IEEE 1609.2 標準選擇 NIST P-256 作為主要橢圓曲線之一，所以 $x$ 坐標和 $y$ 坐標的長度都為 32 bytes [5]。

未壓縮橢圓曲線點(即 uncompressedP256 類型)的結構包含 CHOICE、$x$ 坐標和 $y$ 坐標。已壓縮橢圓曲線點(如：x-only、compressed-y-0、以及 compressed-y-1 等類型)的結構包含 CHOICE 和 $x$ 坐標。每一種類型的結構如表三所示。因此，未壓縮橢圓曲線點的長度(即 $L_1$)可以由公式(11)測量，其中包括 CHOICE 的長度(即 $l_1$)、$x$ 坐標的長度(即 $l_2$) 和 $y$ 坐標的長度(即 $l_3 = l_2$)。此外，已壓縮橢圓曲線點的長度(即 $L_2$)可以由公式(12)測量，其中包括 CHOICE 的長度和 $x$ 坐標的長度。

$$L_1 = l_1 + l_2 + l_3 = l_1 + 2 \times l_2. \tag{11}$$
$$L_2 = l_1 + l_2. \tag{12}$$

表三: 未壓縮橢圓曲線點和已壓縮橢圓曲線點的結構

| 是否壓縮 | 類型 | CHOICE | $x$ 坐標 | $y$ 坐標 |
|---|---|---|---|---|
| 未壓縮橢圓曲線點 | uncompressedP256 | × | × | × |
| 已壓縮橢圓曲線點 | x-only | × | × | |
| 已壓縮橢圓曲線點 | compressed-y-0 | × | × | |
| 已壓縮橢圓曲線點 | compressed-y-1 | × | × | |

在 IEEE 1609.2 標準中，NIST P-256 被選為主要橢圓曲線之一[5]，所以在該橢圓曲線基礎的未壓縮橢圓曲線點和已壓縮橢圓曲線點的長度分別為 65 bytes 和 33 bytes。此外，本研究還使用公式(11)和(12)在不同橢圓曲線[9]下檢查未壓縮橢圓曲線點和已壓縮橢圓曲線點的長度，如表四所示。例如，在 NIST P-521 下，未壓縮橢圓曲線點和已壓縮橢圓曲線點的長度分別為 133 bytes 和 67 bytes，使用已壓縮橢圓曲線點可以節省 66 bytes 的空間。

表四: 未壓縮橢圓曲線點和已壓縮橢圓曲線點的長度(單位：bytes)

| 安全強度 | 橢圓曲線 | 未壓縮橢圓曲線點 | 已壓縮橢圓曲線點 |
|---|---|---|---|
| 80 | NIST P-192 | 49 | 25 |
| 112 | NIST P-224 | 57 | 29 |
| 128 | NIST P-256 | 65 | 33 |
| 192 | NIST P-384 | 97 | 49 |
| 256 | NIST P-521 | 133 | 67 |

## 4.2、已壓縮橢圓曲線點的應用

儘管已壓縮橢圓曲線點可能具有較短的長度，但某些類型的已壓縮橢圓曲線點不能用作公鑰或重構值(Reconstruction Value)。儘管可以使用公式(1)從 $x$ 坐標計算 $y$ 坐標，但因為 $y$ 為平方，可能會生成兩個潛在的 $y$ 坐標。因此，x-only 類型無法還原為未壓縮

橢圓曲線點。然而，在 compressed-y-0 或 compressed-y-1 類型下，可以使用公式(1)，並且結合無窮遠點 $O$ 和(13)計算 $y$ 坐標。此外，根據 compressed-y-0 或 compressed-y-1 的 CHOICE 可以確定橢圓曲線點是$(x, y)$或是$(x, -y)$。

$$(x, -y) = O - (x, y). \tag{13}$$

基於橢圓曲線數位簽章演算法的簽章產製和簽章驗證僅使用 $x$ 坐標(即第 3 節中的 $R_x$)，而從不使用 $y$ 坐標(即第 3 節中的 $R_y$)。因此，x-only、compressed-y-0、compressed-y-1、以及 uncompressedP256 等類型可以用於數位簽章。

對於公鑰和重構值的表示，完整的橢圓曲線點需要包括 $x$ 坐標和 $y$ 坐標。例如，在公式(4)中應指定完整的橢圓曲線點 $P$ 用於金鑰擴展(即重構值)。因此，compressed-y-0 或 compressed-y-1 類型可以用於公鑰和重構值，而 x-only 類型不適用於這些應用。

## 五、評估 ECQV 隱式憑證

在 IEEE 1609.2.1 標準中，採用 ECQV 隱式憑證搭配蝴蝶金鑰擴展機制，節省憑證長度和達成匿名性[6]。5.1 節說明 ECQV 隱式憑證產製流程和主要步驟。5.2 節比較顯式憑證和隱式憑證的結構。最後，本研究評估 ECQV 隱式憑證的潛在威脅，並且提出數學模型證明 ECQV 被破解的概率。

### 5.1、ECQV 隱式憑證產製流程

本節概述 ECQV 隱式憑證產製流程，包括初始階段、憑證請求階段、私鑰擴展階段、以及公鑰擴展階段，如圖一所示[2]-[4]。

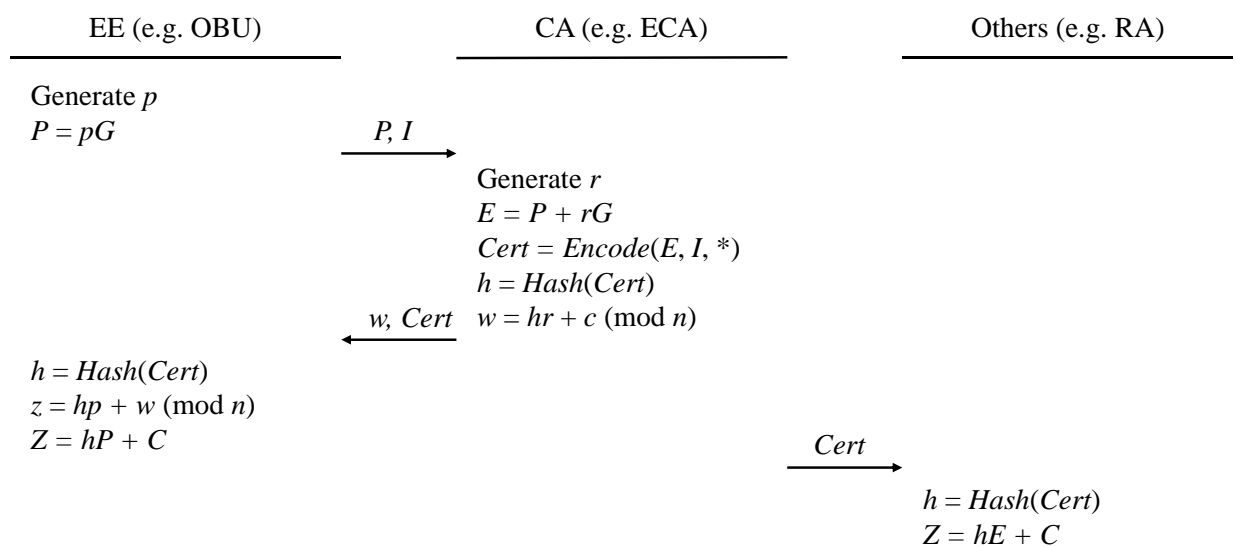

圖一: ECQV 隱式憑證產製流程

在初始階段，終端設備(如：車載設備(On-Board Unit, OBU))可以基於公式(2)產製其 ECC 私鑰 $p$ 和 ECC 公鑰 $P$。此外，憑證中心(Certificate Authority, CA)(如：註冊憑證中心(Enrollment Certificate Authority, ECA))具有其 ECC 私鑰 $c$ 和 ECC 公鑰 $C$。

在憑證請求階段，車載設備可以向註冊憑證中心發送一個註冊憑證(Enrollment Certificate, EC)請求，請求內容包含車載設備公鑰 $P$ 和資訊 $I$。當註冊憑證中心接收到請求時，將產製隨機整數 $r$，並根據公式(4)獲得重構值 $E$。之後，可以產製註冊憑證($Cert$)，憑證內容包含重構值 $E$、車載設備資訊 $I$ 和其他必要的資訊。此外，註冊憑證雜湊(hash)值 $h$ 可以根據公式(14)產製重構私鑰 $w$。最後，將重構私鑰 $w$ 和註冊憑證發送給車載設

備。

$w = hr + c \pmod{n}$ (14)

在私鑰擴展階段，車載設備接收到重構私鑰 $w$ 和註冊憑證($Cert$)，可以計算出註冊憑證($Cert$)的雜湊(hash)值 $h$。根據 $h$、$p$ 和 $w$ 的值，可以使用公式(15)產製擴展私鑰。

$z = hp + w \pmod{n}$ (15)

在公鑰擴展階段，其他設備(如：登記中心(Registration Authority, RA))可以根據註冊憑證中心和車載設備的憑證產製車載設備擴展公鑰 $Z$。基於註冊憑證($Cert$)的雜湊(hash)值 $h$、車載設備憑證中的重構值 $E$、以及註冊憑證中心憑證中的公鑰 $C$。

$Z = hE + C$ (16)

## 5.2、顯式憑證和隱式憑證結構

在 IEEE 1609.2 和 1609.2.1 標準中，顯式憑證中包含版本、類型、簽署者、待簽署憑證(To-Be-Signed-Certificate, TBSC)、以及簽章。待簽署憑證中的驗證金鑰指示器(Verify Key Indicator, VKI)包含公鑰[5]-[6]。此外，顯式憑證需要簽章。因此，顯式憑證的長度($L_3$)可以用公式(17)表示，其中包含已壓縮橢圓曲線點型式的公鑰長度($L_2$)、以及簽章(即帶有已壓縮橢圓曲線點($L_2$)和整數($l_2$)的集合($R$, $s$))，以及其他資訊的長度($q$)。

$L_3 = q + L_2 + L_2 + l_2 = q + 2 \times L_2 + l_2.$ (17)

對於隱式憑證，不需要簽章以減少憑證的長度。此外，待簽署憑證中的驗證金鑰指示器包括含重構值(即圖一中的 $E$) [5]-[6]。因此，隱式憑證長度($L_4$)可以用公式(18)表示，其中包含已壓縮橢圓曲線點型式的重構值長度($L_2$)，以及其他資訊的長度($q$)。

$L_4 = q + L_2.$ (18)

為了比較顯式憑證和隱式憑證的長度，考慮了美國國家標準與技術研究院定義的各種安全強度，並在表五中呈現在不同安全強度下的比較結果。例如，在 NIST P-256 下，顯式憑證的長度為 $q + 98$ bytes，而在相同橢圓曲線下，隱式憑證的長度為 $q + 33$ bytes，減少了 65 bytes。表五顯示在各種橢圓曲線下使用隱式憑證方式減少的長度。

表五: 顯式憑證和隱式憑證的長度(單位：bytes)

| 安全強度 | 橢圓曲線 | 顯式憑證 | 隱式憑證 |
|---|---|---|---|
| 80 | NIST P-192 | $q + 74$ | $q + 25$ |
| 112 | NIST P-224 | $q + 86$ | $q + 29$ |
| 128 | NIST P-256 | $q + 98$ | $q + 33$ |
| 192 | NIST P-384 | $q + 146$ | $q + 49$ |
| 256 | NIST P-521 | $q + 200$ | $q + 67$ |

## 5.3、ECQV 被破解機率

儘管使用 ECQV 可以實現較短的憑證長度，但本節分析 ECQV 相關的潛在威脅。由於重構私鑰 $w$ 包含憑證中心私鑰 $c$，如果滿足某些條件(即 $hr + c < n$)並通過模數公開信息 $h$，則存在憑證中心私鑰 $c$ 可能被快速獲得的風險。因此，本研究提出數學模型，如公式(19)，來評估 ECQV 被破解的機率(即條件 $hr + c < n$ 的機率)。

數學模型假設如下：
· 整數 $h$ 在範圍[1, $H$]內。
· 整數 $r$ 在範圍[1, $R$]內。
· 變數 $c$ 為正整數。

- 變數 $n$ 為質數。
- 變數 $c$ 的值小於 $n$。

$$Pr(H,R)$$
$$= Pr(hr < n-c)$$
$$= \begin{cases} \sum_{h=1}^{n-c} \sum_{r=1}^{\lfloor \frac{n-c}{h} \rfloor} \frac{1}{HR}, \text{if } H > n-c \text{ and } R > \lfloor \frac{n-c}{h} \rfloor, \\ \sum_{h=1}^{n-c} \sum_{r=1}^{R} \frac{1}{HR}, \text{if } H > n-c \text{ and } R \leq \lfloor \frac{n-c}{h} \rfloor, \\ \sum_{h=1}^{H} \sum_{r=1}^{\lfloor \frac{n-c}{h} \rfloor} \frac{1}{HR}, \text{if } H \leq n-c \text{ and } R > \lfloor \frac{n-c}{h} \rfloor, \\ \sum_{h=1}^{H} \sum_{r=1}^{R} \frac{1}{HR}, \text{if } H \leq n-c \text{ and } R \leq \lfloor \frac{n-c}{h} \rfloor. \end{cases} \quad (19)$$

$$= \begin{cases} \frac{(n-c)\lfloor \frac{n-c}{h} \rfloor}{HR}, \text{if } H > n-c \text{ and } R > \lfloor \frac{n-c}{h} \rfloor, \\ \frac{n-c}{H}, \text{if } H > n-c \text{ and } R \leq \lfloor \frac{n-c}{h} \rfloor, \\ \frac{\lfloor \frac{n-c}{h} \rfloor}{R}, \text{if } H \leq n-c \text{ and } R > \lfloor \frac{n-c}{h} \rfloor, \\ 1, \text{if } H \leq n-c \text{ and } R \leq \lfloor \frac{n-c}{h} \rfloor. \end{cases}$$

當整數 $H$ 小於 $n$ - $c$ 且整數 $R$ 小於 $\lfloor \frac{n-c}{h} \rfloor$ 時，可能破解出憑證中心私鑰 $c$。

## 六、討論與建議

基於從第 2 節到第 5 節的分析，本研究針對安全憑證管理系統效能改進和安全性提出以下建議：

- 金鑰產製和金鑰擴展的計算時間主要花費在計算公鑰的橢圓曲線點。因此，產製的公鑰和擴展的公鑰可以存儲在晶片裡，以減少重覆的計算時間。
- 簽章產製和簽章驗證的計算時間主要花費在計算橢圓曲線點 $R$ (即 $R = rG$)。因此，可以設計安全的偽隨機數產生器來產製安全且特別的隨機整數，從而減少計算時間。
- 已壓縮橢圓曲線點是減少簽章和憑證長度的有效方法。但是，x-only 類型無法用作公鑰的橢圓曲線點。因此，compressed-y-0 或 compressed-y-1 更適合通用於數位簽章和公鑰的已壓縮橢圓曲線點。
- ECQV 是實現較短憑證長度和基於蝴蝶金鑰擴展機制的假名憑證的有效方法。然而，在特定條件下(即 $hr + c < n$)，通過模數運算利用公開信息 $h$，則憑證中心的私鑰 $c$ 可能被迅速獲取的潛在風險。為了減輕這種風險，建議產製更大的隨機整數 $r$，以避免 ECQV 被破解的威脅。

## 七、結論與未來研究

本研究評估了金鑰產製、金鑰擴展、簽章產製、以及簽章驗證的計算時間。此外，

本研究還分析了未壓縮橢圓曲線點、已壓縮橢圓曲線點、顯式憑證、以及隱式憑證的長度。基於這些分析，本研究提出了增強和最佳化這些標準化安全技術在 V2X 通訊中效能的建議。此外，本研究指出了 ECQV 被破解的潛在風險，並通過數學模型證明 ECQV 破解的機率。考慮到 ECQV 被破解的可能性，本研究建議在 ECQV 中使用更大的隨機整數，以減輕 ECQV 破解的潛在威脅。

在未來研究中，鑒於量子計算所帶來的潛在威脅，值得考慮在未來 V2X 安全憑證管理系統中結合後量子密碼學方法，提升安全性和抗量子計算攻擊的能力。

## 參考文獻